\documentclass[journal]{article}

\usepackage{graphicx,amsmath,amssymb,overpic,here,multirow,listings, bm, hyperref, here}        

\usepackage[dvipsnames]{xcolor}
\usepackage{tikz}
\usepackage{verbatim}
\usetikzlibrary{arrows,shapes}
\usepackage{pgfplots}
\usetikzlibrary{positioning}
\usepackage{framed}
\usepackage{mdframed}

\definecolor{shadecolor}{RGB}{180,180,180}

\usepackage{algorithm}
\usepackage{algpseudocode}

\begin{document}
%
\title{On the use of Data-Driven Cost Function Identification in Parametrized NMPC}

%
%
%

\author{Mazen Alamir \thanks{The material in this paper was not presented at any conference. M. Alamir is with Univ. Grenoble Alpes, CNRS, Grenoble INP*, GIPSA-lab, 38000 Grenoble, France. Email: mazen.alamir@grenoble-inp.fr  (http://www.mazenalamir.fr). This work has been partially supported by MIAI @ Grenoble Alpes, (ANR-19-P3IA-0003)}
}

%
%

\markboth{Submitted to IEEE Trans. on Automatic Control}%
{Shell \MakeLowercase{\textit{et al.}}: Bare Demo of IEEEtran.cls for Journals}
%



\date{\ }

\maketitle

\begin{abstract}
In this paper, a framework with complete numerical investigation is proposed regarding the feasibility of constrained Nonlinear Model Predictive Control (NMPC) design using Data-Driven model of the cost function. Although the idea is very much in the air, this paper proposes a complete implementation using python modules that are made freely available on a GitHub repository. Moreover, a discussion regarding the different ways of deriving control via data-driven modeling is proposed that can be of interest to practitioners. 
\end{abstract}


%

\section{Introduction}
\noindent  While in the very heart of the standard traditional identification schemes from the early years of automatic control, the use of  data-driven models in control design witnessed a renewable surge these last years due to the emergence of user-friendly tools of {\em map fitting} libraries such as \lstinline{scikitlearn} \cite{scikitlearn} and \lstinline{TensorFlow} \cite{tensorflow2015} to cite but few ones. These tools enable to try, in an astonishingly effortless way, several possible structures in order to map static relationships between a vector of inputs (the regressor) and some targeted variable (label) to be identified. 
\\ \ \\ 
When it comes to their use in solving standard control problems, the emergence of these tools in the control community induces quite often very passionate debates regarding their true novelty when compared to the traditional ways of handling the same problems in the control community. As soon as nonlinear systems are concerned, the opportunities offered by these tools should be, and is in fact, less debatable as they clearly enable an easy and rapid prototyping of candidate solutions to many nonlinear control problems. 
\\ \ \\ 
Regarding the use of nonlinear fitting maps in control design, several approaches can be used depending on the choice of the target variable for which a model is searched for. Indeed:
\\ \ \\ 
{\bf A first option} is to use the optimal solution (the first control in the optimal sequence over the prediction horizon) as a target variable. This is particularly appealing since it avoids the need to do on-line optimization as it is typically done in standard Model Predictive Control design. For this to be done, a dynamic model of the controlled system is needed. This model can therefore be used to solve off-line a high number of open-loop optimal control problems so that a learning data can be obtained for the derivation of the model via nonlinear regression problem settings. Each of these problems is obtained by choosing a different initial state of the presumably available dynamic model. Beside the need for the complete model for the formulation of the open-loop optimal control problems, the computation time is expected to be quite important as for each sample in the learning data, a constrained NLP problem need to be solved off-line using some appropriate NLP solver. Finally, even if a mathematical model is available, the values of its parameters are generally not well known and all mismatches in these values are inherited by the learning data leading to detuned results that have to be corrected in some way. 
\\ \ \\ 
{\bf A second option} might be used when the equations of the dynamics are not available. In this case, the data-driven tools might be used to derive an input/output nonlinear model of the dynamics. The latter can then be used in two ways:
\\ \ \\ 
	a) Either as in the previous approach in which the learned model is used to define a high number of optimal control problems to be solved off-line. 
\\ \ \\ 
	b) Or the model is used in a standard on-line NMPC framework where repetitive on-line solution of the underlying sequence of optimal control problems is solved using fast NMPC approaches. 
\\ \ \\ 
	In this option, the original task is rather ambitious because one is seeking a complete faithful model of the whole dynamics of the system that can is generally very difficult to achieve even for almost linear systems. 
\\ \ \\ 
{\bf The third option} concerns also the case where no mathematical model is available but here, one does not try to identify a simulator of the system. Rather, one tries to identify the relationship between past measurements, past and future control {\em on one hand} and  a resulting cost value that is defined on the future prediction horizon {\em on the other hand}. In order to get the data that makes the identification of such map possible, it should be assumed that some initial loosely defined control (that might be open-loop or roughly defined output feedback) can be applied to the system over a sufficiently long time in order to gather a rich set of measurement data for the above-mentioned identification task. This requirement is satisfied for a reasonably large class of systems. Note that the resulting model does not allow system simulation as in the last option, but it hopefully captures the precisely needed relationship that enables the underlying optimal control problem to be defined at each decision instant based on the previously obtained measurement sequences. 
\\ \ \\ 
The present contribution follows this third option and proposes a complete framework that is tested on an illustrative example. Moreover, all the tools used in the implementation of the proposed solution, namely the preprocessing of the data, the identification of the model and the solution of the resulting optimization problems, are made available on an open GitHub repository\footnote{https://github.com/mazenalamir/siso\_predictor, https://github.com/mazenalamir/torczon} \cite{mizorepo2020sisopredictor}. 
\\ \ \\ 
The use of the newly available Machine Learning libraries in addressing many control problems is a very wide investigation area and it is unlikely that any additional contribution can legitimately claim a total novelty as far as general ideas are concerned. However, as the {\em Devil is in the details}, any complete exposition of a successful application of such ideas together with the availability of the tools and lines of code that made this success possible can be of great help to many researchers that are seeking concrete and accessible recipes. This is the aim of the present contribution. \\ \ \\ 
This paper is organized as follows. First, some definitions and notation are stated in Section \ref{sec-def-not}. The illustrative example that is used throughout the paper to support the presentation is explained in Section \ref{sec-illustrative-example}. The proposed framework is detailed in Section \ref{sec-proposed-framework} while the results are shown in Section \ref{sec-res} together with an extended discussion. Finally Section \ref{sec-conc}.

\section{Definition and Notation} \label{sec-def-not}
\begin{figure}[H]
\begin{center}
\includegraphics[width=0.45\textwidth]{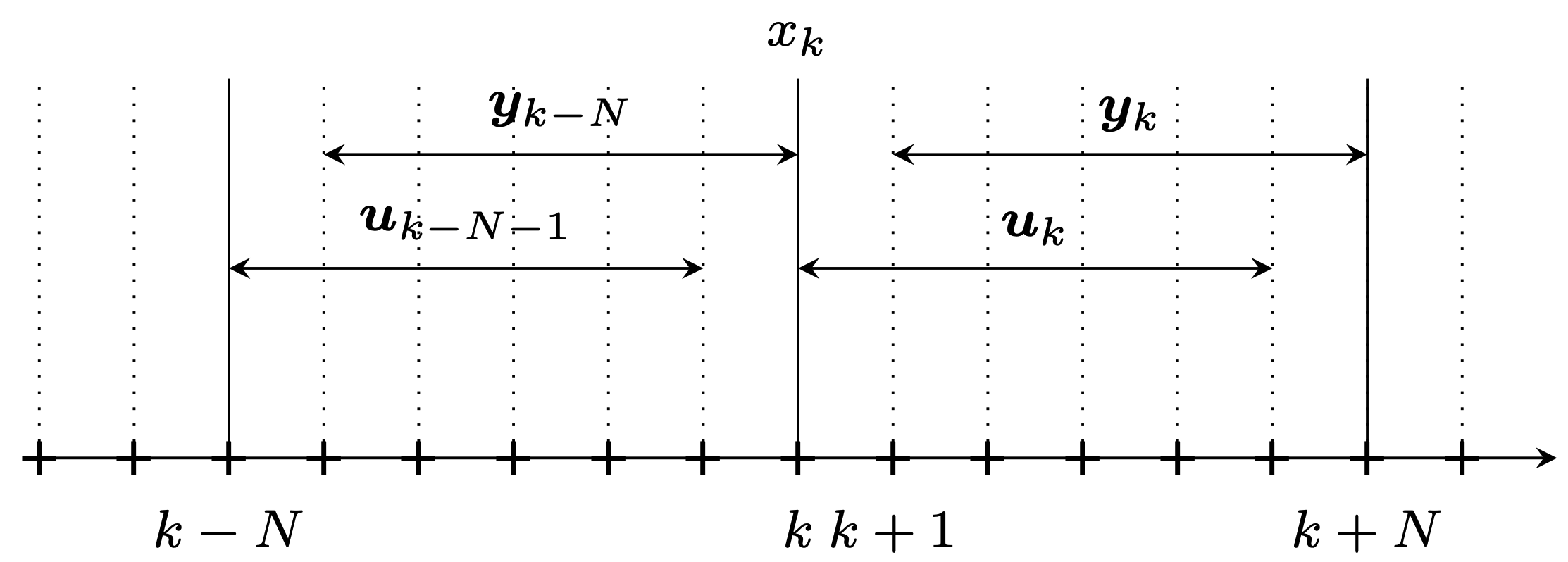}
\end{center}	
\caption{Illustration of the observability property: $(\bm y_{k-N}, \bm u_{k-N-1})$ uniquely determine $x_k$ which, together with $\bm u_k$ uniquely determine $\bm y_k$.} \label{fig-observability}
\end{figure}

Consider a nonlinear system to be controlled that can be described by some {\bf  unknown dynamics} given by
\begin{equation}
x^+=f(x,u)\qquad y=h(x,u) \label{lesyst}
\end{equation}
where $x\in \mathbb{R}^{n_x}$, $u\in \mathbb{R}^{n_u}$ and $y\in \mathbb{R}^{n_y}$ stand for the state vector, the control input vector and the vector of measured variables respectively. Since the maps $f$, $h$ and the very definition of the state vector $x$ are all supposed to be unknown, it is assumed that a cost function expressing the quality of the system's trajectories over some prediction horizon can be expressed, at each instant $k$, through an expression cost$(\bm y_k, \bm u_k)$ that involves the only measured quantities, namely $\bm y_k$ and $\bm u_k$ where:
\begin{equation*}
\bm y_k := \begin{bmatrix}
 y_{k+1}&\dots& y_{k+N}
\end{bmatrix} \in [\mathbb{R}^{n_y}]^{N}
\end{equation*}
\begin{equation*}
\bm u_k := \begin{bmatrix}
 u_{k}&\dots&u_{k+N-1}
\end{bmatrix} \in [\mathbb{R}^{n_u}]^N
\end{equation*}
The feasibility of any output feedback design is based on the observability assumption according to which there is some static map $\mathcal O$ such that for sufficiently high $N$, one has:
\begin{equation}
\bm y_k=\mathcal O(\bm y_{k-N}, \bm u_{k-N-1}, \bm u_k) \label{obs}
\end{equation}
as this simply stems from the fact that the state $x_k$ is inferred from $(\bm y_{k-N}, \bm u_{k-N-1})$ which together with $\bm u_k$ determine the future $\bm y_k$ (see Figure \ref{fig-observability}). 
\ \\ 
A direct consequence of the observability property is that the cost function cost$(\bm y_k, \bm u_k)$ can be written in the following form cost$(\mathcal O(\bm y_{k-N}, \bm u_{k-N-1}), \bm u_k)$ or equivalently, using a straightforward definition of $J$:
\begin{align}
\mbox{\rm cost}&(\ \overbrace{\bm y_{k-N}, \bm u_{k-N-1}}^{\mbox{\color{gray} \footnotesize past measurement}}, \overbrace{ \bm u_k}^{\mbox{\color{gray} \footnotesize Future action}}) \label{defdecost}\\ 
&=: J(\bm u_k\ \vert \ \bm y_{k-N}, \bm u_{k-N-1}) \label{defdeJ}
\end{align}
where (\ref{defdeJ}) is a simple rewriting of (\ref{defdecost}) after reordering the terms so that $J$ can be identified as:
\\ \ 
\begin{center}
\begin{minipage}{0.48\textwidth}
\begin{mdframed}[backgroundcolor=Gray!20, linecolor=white]
{\em The cost induced by a future control sequence $\bm u_k$ given the past measurements $(\bm y_{k-N}, \bm u_{k-N-1})$}
\end{mdframed}
\end{minipage}
\end{center}
 \ \\ 
This cost can then obviously be used to design NMPC output feedback by solving at each instant $k$ the optimization problem:
\begin{equation}
\bm u_k^*\leftarrow {\rm arg}\min_{\bm u\in \mathbb U^N} J(\bm u\ \vert  \ \bm y_{k-N}, \bm u_{k-N-1}) \label{defdeoptimpb}
\end{equation} 
and by applying the first control $u_k^*$ in the so-computed optimal sequence $\bm u_k^*$ (see \cite{Mayne2000} for more details regarding MPC design). 
Note that in (\ref{defdeoptimpb}) the set $\mathbb U\subset \mathbb{R}^{n_u}$ stands for the admissible domain for the input vector. This implements box-like hard constraints on the control. Should output constraints be incorporated, they can appear in the definition of the cost function through constraints penalty leading to soft implementation of such constraints. These obvious details are omitted here for the sake of conciseness. \\ \ \\  In the next section, an illustrative example is introduced so that the concepts can be associated to a concrete use-case. This example is also used in the numerical investigation-related section. 
\section{Illustrative Example} \label{sec-illustrative-example}
Consider the example of the nonlinear continuous flow stirred-tank reactor with parallel reactions \cite{Bailey1971}:
$$R\rightarrow P_1\qquad R\rightarrow P_2$$
This system is commonly modeled by the following set of highly nonlinear Ordinary Differential Equations (ODEs):
\begin{subequations}
	\begin{align}
\dot x_1&=1-10^4x_1^2e^{-1/x_3}-400x_1e^{-0.55/x_3}-x_1 \label{reac1}\\
\dot x_2&=10^4x_1^2e^{-1/x_3}-x_2 \label{reac2}\\
\dot x_3&=u-x_3 \label{reac3}
\end{align}
\end{subequations}
where $x_1$ and $x_2$ stand for the concentrations of $R$ and $P_1$ respectively while $x_3$ represents the temperature of the mixture in the reactor. $P_2$ represents the waste product. The control variable is given by the heat flow $u\in \mathbb U:=[0.049, 0.449]$. The objective of the control is to maximize the amount of product $x_2=P_1$ which is precisely the only measured variable. It can be checked that this output makes the system completely observable. \\ \ \\ 
This system is recurrently invoked in the works related to the so-called {\bf Economic MPC} \cite{Muller2014} where the cost function is not based on a priori known steady state. Instead, only an economic stage cost $\ell(x)=-x_2=-y$ is used to enhance the maximization of the production. More precisely, the following structure of the cost function is typically used:
\begin{equation}
-\bigl(\dfrac{1}{N}\mathbf{1}^T\cdot \bm y_k+\alpha y_{k+N}\bigr) \label{lacout}
\end{equation}
where the first term is noting but the average value over the prediction horizon while the second terms implements a terminal penalty at the end of the prediction horizon following standard stability-related recommendations \cite{Mayne2000}. The minus sign recalls that maximization is required. \\ \ \\ 
The following are some known facts that are relevant for a better understanding of the numerical results shown in the remainder of this paper:
\begin{enumerate}
\item The consequence of the EMPC formulation is that optimal trajectories are not necessarily steady but rather oscillating around the optimal steady state since these oscillations potentially induce a larger production. \\
\item The optimal solution is very sensitive to the initial state. This can be felt when using even sophisticated solvers \cite{Andersson2018} in  the fact that they need rather high numbers of iterations even when warm start is used during the closed-loop simulations. This means that the underlying optimization problems are rather challenging. This should  be kept in mind when appreciating the quality of the results under the identified maps. 
\end{enumerate}
In the forthcoming developments, a sampling period of $\tau=0.02$ time units is used (note that the equations above are normalized so that the absolute time is meaningless). Within all possible steady pairs, the optimal one corresponds to $x_{st}=(0.0832, 0.0846, 0.149)$ and $u_{st}=0.146$ and hence $y_{st}=0.0846$.   
\section{Proposed Framework} \label{sec-proposed-framework}
The key tasks in the implementation of the output NMPC feedback as sketched in the previous section are the following:
\begin{enumerate}
\item Define an initial input choice (initial output controller or simple open-loop) to generate the learning data (Section \ref{sub-generate-data}).\\
\item Identify the map (\ref{defdeJ}) using the previously generated learning data and an appropriately chosen model's structure (Section \ref{sub-identification}).\\
\item Solve the constrained optimization problem (\ref{defdeoptimpb}) at each decision instant using a dedicated NLP solver (Section \ref{sub-solve}). 
\end{enumerate}
These steps are successively detailed in the following subsections. Obviously there are many possible choices for each of the above mentioned tasks. The framework sketched here proposes a set of choices and examine how the resulting framework behaves in closed-loop. 
\subsection{Generate the learning data} \label{sub-generate-data}
\begin{figure}[H]
\begin{center}
\includegraphics[width=8.5cm, height=4cm]{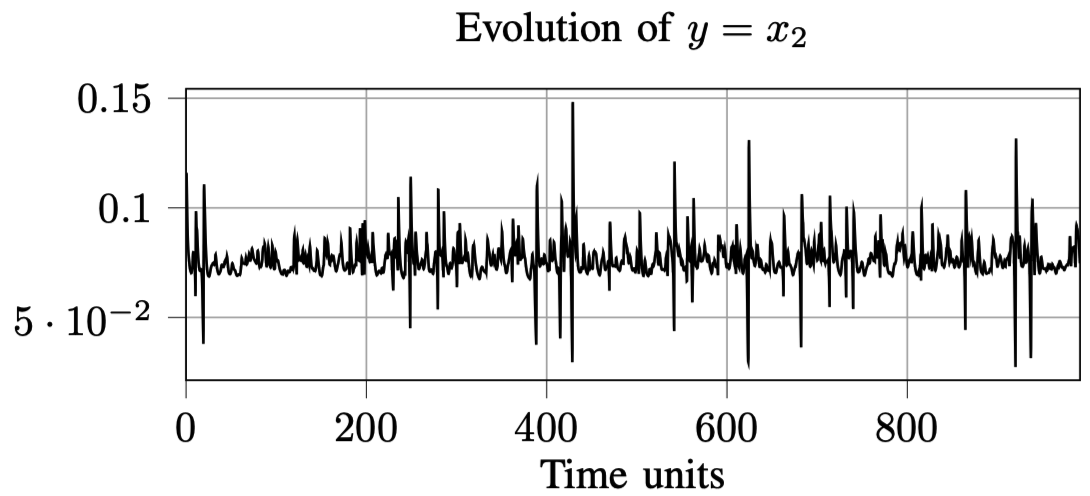}
\end{center}	
\caption{Output evolution under the random input sequence shown in figure \ref{input_ol}. This time series contains 50324 instants.} \label{output_ol}
\end{figure}

\begin{figure}[H]
\begin{center}
\includegraphics[width=8cm, height=4cm]{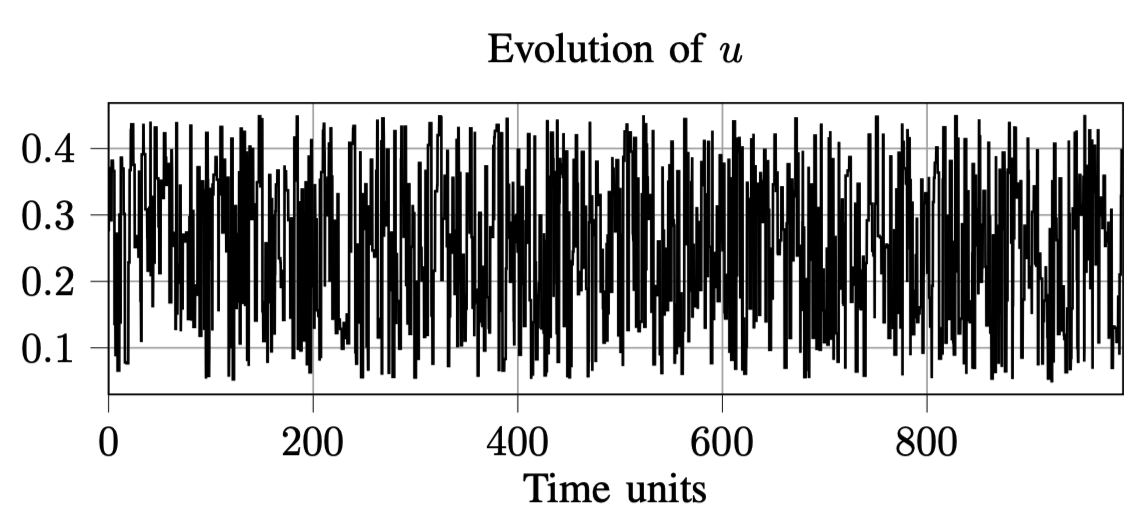}
\end{center}	
\caption{Randomly sampled input sequence in $\mathbb U:=[0.049, 0.449]$.} \label{input_ol}
\end{figure}

This task is probably the most problem-dependent among the above enumerated tasks. In some cases, sub-optimal explicit initial output feedback can be implemented, for others, simple random sequences can be applied. In the case of the example under consideration, a long random sequence of input is sampled uniformly inside the admissible set $\mathbb U:=[0.049, 0.449]$. This sequence is applied to the system and the resulting output is collected for the learning phase. Figures \ref{output_ol}-\ref{input_ol} show typical resulting sequences of input/output measurements. 

To be more precise regarding the way the input signal is generated, the following procedure is used to induce explicitly non uniform piece-wise time segment lengths:\\
\begin{itemize}
\item First a set of $1000$ random values of interval durations is sampled uniformly in the set of integer multiple of $\tau$ lying inside $[\tau,100\tau]$. This leads to durations randomly sampled between $0.02$ and $2$ time units. \\
\item A corresponding set of $1000$ values of the input uniformly distributed inside $\mathcal U:=[0.049, 0.149]$ has been sampled. 
\end{itemize}
Figures \ref{output_ol}-\ref{input_ol} show one realization that resulted from the above procedure. Experience showed that the overall results in terms of closed-loop performances are quite insensitive t to the specific realization that is used to learn the model provided that the duration is sufficient to enable the learning data to be representative. 
\subsection{Identification of the cost function} \label{sub-identification}
\begin{figure}
\begin{center}
\includegraphics[width=0.5\textwidth]{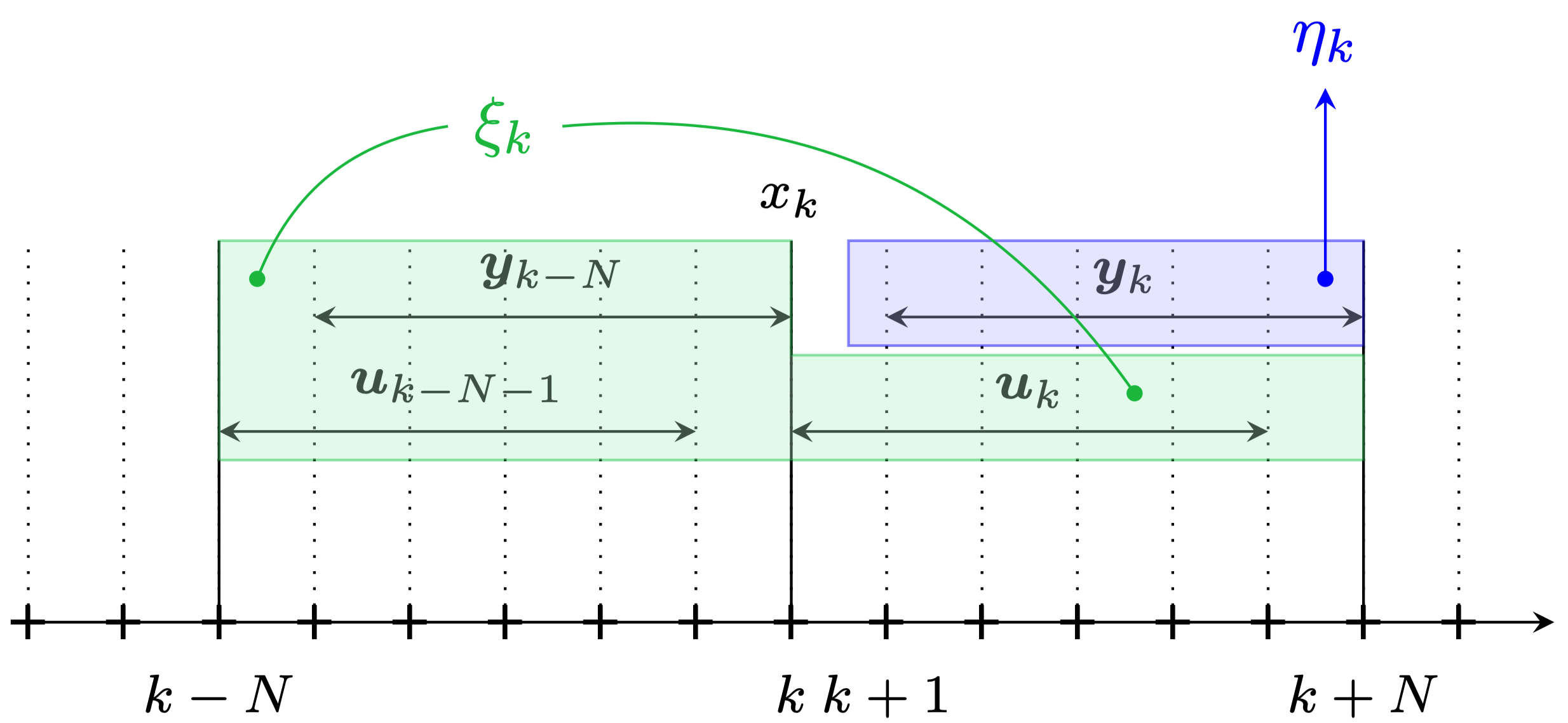}
\end{center}	
\caption{The learning data is built by a moving window over the generated open-loop trajectory. Definition of $\xi_k$ 	and $\eta_k$. Using a rolling windows of length $N$ over the time series depicted on Figures \ref{output_ol}-\ref{input_ol} that contains 50324 lines, it is possible to extract a number of $n_r=50324-2N$ samples to build the data set $\mathcal D$} \label{fig-map}
\end{figure}

The above generated data set is used here to identify the relationship (\ref{defdecost}) [or equivalently (\ref{defdeJ})]. This problem is a standard {\bf  regression} problem where one seeks a function $F$ that maps a regressor $\xi$ to some continuous label $\eta$, namely:
\begin{equation}
\mbox{\rm (Regression problem) \hskip 3mm Find $F$ s.t} \quad F(\xi)\approx \eta
\end{equation}
the $\approx$ is to be understood in terms of some measure of the statistics of the errors $\eta_i-F(\xi_i)$ over a set of realizations $\mathcal D:=\{(\xi_i,\eta_i)\}_{i=1}^{n_r}$ that is called the learning set. Note that in the case of the example under study, the regressor and the label are defined by (see Figure \ref{fig-map}):
\begin{align}
\xi &=(\bm y_{k-N}, \bm u_{k-N-1}, \bm u_k) \label{defdexi} \\
\eta &= J(\bm u_k\ \vert\ \bm y_{k-N}, \bm u_{k-N-1}) \label{defdeeta}
\end{align}
Using a rolling windows of length $N$ over the time series depicted on Figures \ref{output_ol}-\ref{input_ol} that contains 50324 lines, it is possible to extract a number of $n_r=50324-2N$ samples to build the data set $\mathcal D$ mentioned above. \\ \ \\ 
The quality of the identified map is generally evaluated using a so called test set that contains realizations that are not  present in the learning set in order to check the generalization power of the identified map (or to say it in other words, in order to evaluate the degree of overfitting that might occur when using a too complex model compared to the available data).
As it is shown later in the validation section, the window size $N$ can be as big as $100$ when using cost functions that are defined on a prediction horizon of 2 time units (recall that the sampling period is $\tau=0.02$ which is needed to ensure stability of the numerical integration as the system dynamics is stiff. Therefore, when considering the regressor $\xi$ defined by (\ref{defdexi}), this would lead to a regressor of dimension $300$ which is obviously prone to overfitting (too many parameters). \\ \ \\ 
A typical approach to reduce the size of the regressor is to use Principal Component Analysis feature reduction method. However when it comes to handle physical time series, this approach is not necessarily the lighter one or even the more appropriate. Rather, it is advised here to use the following {\em moment-like} features consisting in considering the average of the successive derivatives of a smoothed version of the signals over the time intervals. More precisely, given a signal $s$ defined on some interval $I$,  we define the following features:
\begin{equation}
\phi_j(s) := \dfrac{1}{\vert I\vert }\sum_{i=1}^{\vert I\vert} s^{(j)}(i) \quad j\in \{0,1,\dots,m-1\}\label{defdephij}
\end{equation}
where $s^{(j)}(i)$ is the value at instant $i$ of the  $j$-th derivative of the signal $s$ with the convention $s^{(0)}\equiv s$. This feature extractor is implemented in the utility {{\color{Blue} \lstinline{feature_gen}} of the module  \lstinline{siso_predictor} available on \href{https://github.com/mazenalamir/siso_predictor}{the GitHub repository} \cite{mizorepo2020sisopredictor}. \\ \ \\ 
Therefore by extracting the $m$ features of $\bm y_{k-N}$, $\bm u_{k-N}$ and $\bm u_k$ that are involved in the definition (\ref{defdexi}), one defines the following features extraction map:
\begin{align}
\Phi(\xi):= \Bigl[&
 \phi_0(\bm y_{k-N}),\dots, \phi_{m-1}(\bm y_{k-N}), \nonumber \\
 &\phi_0(\bm u_{k-N}),\dots, \phi_{m-1}(\bm u_{k-N}), \nonumber \\
 &\phi_0(\bm u_k),\dots, \phi_{m-1}(\bm u_k)
\Bigr] \in \mathbb{R}^{3m}
\end{align}
As far as the example is concerned, $m=3$ is used in the implementation leading to a regressor of dimension 9. \\ \ \\ 
Having this reduced vector of features $\Phi(\xi)$ at hand, one can choose one of the many available {\em Machine Learning} nonlinear structures (SVM, Random Forest, Neural Network, Ridge, Lasso, to cite but a few of available structures) in order to solve the regression problem that can be stated shortly denoted as follows:
\begin{equation}
\eta\approx \mbox{ML}(\Phi(\xi))=:\hat J(\xi) \label{defdehatJ}
\end{equation}
As far as the example is concerned, a Random Forest ({\bf RF}) structure with a controlled maximum number of leaves nodes has been used\footnote{https://scikit-learn.org/stable/modules/generated/sklearn.ensemble. RandomForestRegressor.html}. As a matter of fact, the utility {{\color{Blue} \lstinline{learn_model}} of the module  \lstinline{siso_predictor} available on \href{https://github.com/mazenalamir/siso_predictor}{the GitHub repository} \cite{mizorepo2020sisopredictor} enables to built a piece-wise RF model with a desired number of clusters in order to increase the precision of the model while controlling the complexity of each sub-model). An example of the fitting results that can be achieved using a 66-33\% of training/testing split of the data set is shown in Figure \ref{result_diag}. When clustering is required, the clustering is performed based on the $3m$ features contained in $\Phi(\xi)$  using a standard K-Nearest-Neighbors clustering algorithm of the scikitlearn library \cite{scikitlearn}. 

\begin{figure}[H]
\begin{center}
\includegraphics[width=0.4\textwidth]{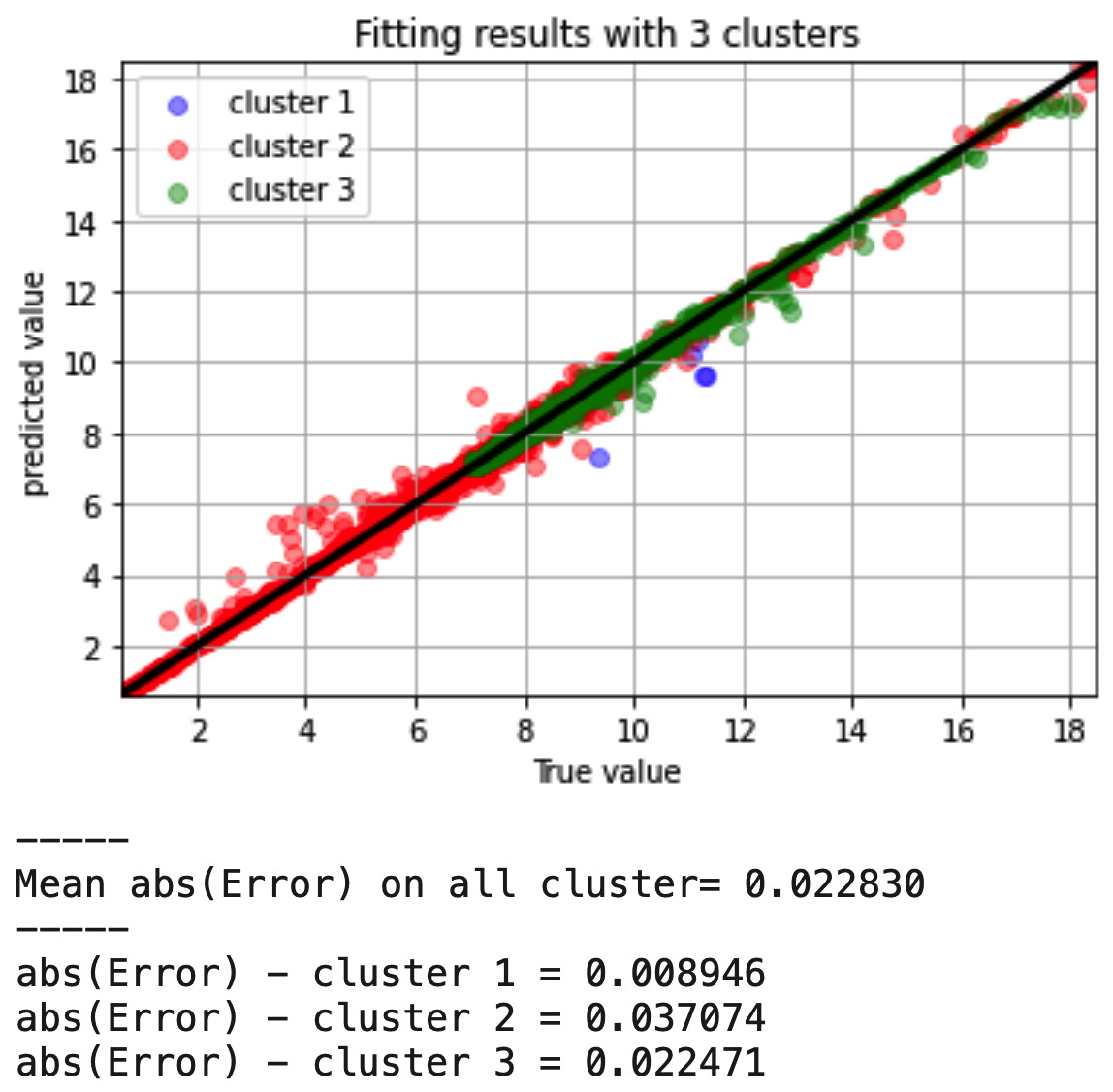}
\end{center}	
\caption{Typical fitting results with statistics on the prediction error. $N=50$ and $\alpha=100$ is used in the terminal penalty (\ref{lacout}).} \label{result_diag}
\end{figure}
\subsection{Solving the optimization problem} \label{sub-solve}
Recall that one is interested in finding the best sequence of future control given the previous measurement of output and control over some past measurement window of length $N$. The resulting optimization problem can therefore be expressed as follows:
\begin{align}
&P(\bm y_{k-N}, \bm u_{k-N}):\nonumber \\
&\bm u^*_k\leftarrow \mbox{arg}\min_{\bm u\in \mathbb U^N} \hat J(\bm y_{k-N}, \bm u_{k-N}, \bm u) \label{lepb}
\end{align}
in which $(\bm y_{k-N}, \bm u_{k-N})$ are the parameters of the problem (obtained from the past window $[k-N,k]$) while the future sequence of control $\bm u_k$ is the decision variable. Note however that the dimension of the decision variable is $(Nn_u)$ which can be huge. We know however that due to the very definition of the feature extraction map $\Phi$ introduced in the previous section:
\begin{center}
\begin{minipage}{0.48\textwidth}
\begin{mdframed}[backgroundcolor=Gray!20, linecolor=white]
All the control profiles that share the same $m$ first moments lead to the same predicted cost function. 
\end{mdframed}
\end{minipage}
\end{center}
 \ \\ 
 \begin{figure}[H]   
\begin{center}
\includegraphics[width=0.5\textwidth]{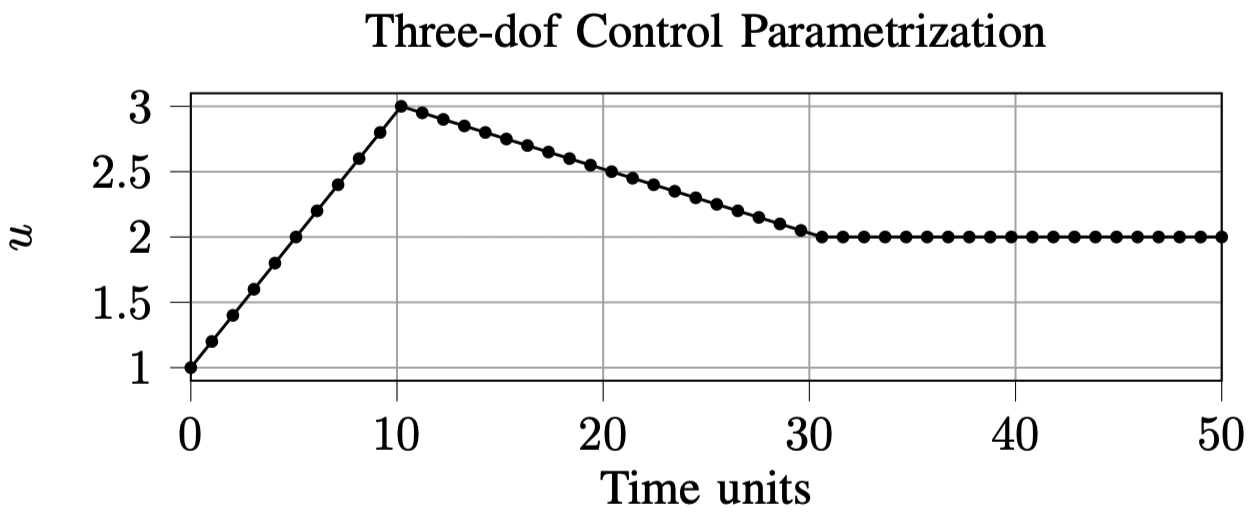}
\end{center}	
\caption{Example of piece-wise linear parameterization of the future control profile using three-degrees of freedom which are the values of the control at instants $k=0, 10$ and $30$. $N=50$ in this example. } \label{fig-uparam}
\end{figure}
\noindent Consequently, there is no benefit from having too many degrees of freedom in the decision variable $\bm u$ involved in (\ref{lepb}) that is the reason why a low-dimensional parameterization of the control profile $\bm u$ is used that leaves as free only the values of the control input at few specific future instants on the prediction horizon. These values induce the remaining values by linear interpolation. Figure \ref{fig-uparam} shows an example of such piece-wise linear control profile over a prediction horizon of length $N=50$ and using  thee d.o.f at instants $k=0,10$ and $30$. \\ \ \\ 
Using this parameterization, one can denote the future control profile by $$\bm u:=\mathcal U(p)$$ where $p$ is the new vector of d.o.f and the optimization problem (\ref{lepb}) to be solved becomes:
\begin{align}
&P_r(\bm y_{k-N}, \bm u_{k-N}):\nonumber \\
&\bm p^*_k\leftarrow \mbox{arg}\min_{\bm p\in \mathbb U^{n_p}} \hat J(\bm y_{k-N}, \bm u_{k-N}, \mathcal U(\bm p)) \label{lepb}
\end{align}
Now one might say that at this stage, it is about solving an optimization problem which can be done using many available solvers. This is only partially true because many of the map structures that can be invoked from the ML libraries to fit a nonlinear function as expressed in (\ref{defdehatJ}) are discontinuous by nature (this is in particular true for the Random Forest regressor). This means that the optimization method should be selected with some care. \\ \ \\ 
To address this issue, it is proposed here to use a modified version of the derivative-free Torczon algorithm \cite{Torczon1991,Torczon1997} that is made available in GitHub repository \cite{mizorepo2020Torczon}. This algorithm is particularly adapted to discontinuous cost functions and it incorporates a multiple starting point mechanism in order to avoid local minima. In a nutshell, the Torczon algorithm starts with a regular polygone in $\mathbb{R}^{n}$ where $n$ is the dimension of the decision variable and improves  it iteratively based on the values of the cost functions at its nodes. The updating involves three possible operations: contraction, reflexion and extension depending on the success/failure of the reflexion beyond the best node. Only function evaluation is required and discontinuity of the cost function induces no specific issue. 

\section{Results and Discussion} \label{sec-res}
The following parameters are used in the closed-loop simulations: 
\begin{itemize}
\item Prediction horizon $N=100$, sampling period $\tau=0.02$
\item The parameterization of the control profile is done using two different options:
\begin{enumerate}
\item Either two degrees of freedom parameterization with the values of the control at the sampling instants $\{0,10\}$ taken as free decision variables to define the piece-wise affine control profile. In this case, one single initial guess ($N_{guess}=1$) is used and the maximum number of iterations of the Torczon algorithm is fixed to $N_{iter}=30$. \\
\item Or Four degrees of freedom at the sampling instants $\{0,5,10,20\}$. Here, $N_{guess}=3$ and $N_{iter}=50$ are used.
\end{enumerate}
$9$ different initial states have been used to check the robustness of the result to context variations. For each of these initial states, the closed-loop system under each of the above settings is simulated. The results are shown in Figures \ref{Res_N100_dof2} and \ref{Res_N100_dof4} hereafter. \\ 
\item The identification of the model has been done using a random forest regressor with a single cluster and a maximum number of leaves nodes set to $1200$.\\
\item Note that since the cost function prediction model needs the measurement over the past window of length $N$, the first $N$ control actions are taken randomly waiting for the buffer to be filled. This corresponds to the first $2$ time units of the simulations. \\
\end{itemize}
It can be noticed that the results are globally quite good in terms of the convergence towards the vicinity of the optimal value $0.0846$. Note however that the oscillations that is the signature of the fully optimal behavior are not present {\bf  probably due to the low dimensional parameterization} that prevents the cycles to take place since its open-loop realization is not inside the possible options offered by the parameterized control profile. By the way, this steady behavior can be the desired one since the slight increase in production induced by the fully optimal limit cycles is obtained at the price of undesirable fatigue on the actuator. \\ \ \\ 
It can also be observed that the quality of closed-loop is slightly better in the case where the second setting is used. However, one must keep in mind that the maximum number of iterations is $5$ times higher in the second option compared to the first. ($3\times 50$ compared to $1\times 30$). In fact what does really matter here is to assess the fact that the identified model seems appropriate and the remaining issue is a matter of optimization parameters tuning which is a typical problem/hardware dependent trade-off.
\subsection{Computation time}
Although the computation time is not the topics of this paper, it is worth giving some hints regarding the complexity of the resulting on-line computation. Computation is done using Python 3 jupyter-notebook on a MacBook-Pro 2017, Mojave 10.14.5 with a 2.9 GHz Core i7 processor. The computation of a single evaluation of the identified Random-Forest cost function (\ref{defdehatJ}) takes less than 1.7 ms. Since the  Torczon algorithm performs  ($n+4$) evaluation per iteration (where $n$ is the number of decision variables), this means that the maximum MPC computation time when $n=2$, $N_{iter}=30$ and $N_{guess}=1$ (the scenario of Figure \ref{Res_N100_dof2})  is about 0.3 s against 1.53 s for the settings of Figure \ref{Res_N100_dof4}. 

\section{Conclusion and Future Work} \label{sec-conc}
In this paper, a complete framework for data-driven control is presented and tested through a rather challenging nonlinear control problem. The framework applies to the class of dynamical system for which it is possible to apply a random or a priori simple output feedback control law which can be far from being optimal but which enable a learning data set to be built. A third obvious condition is that the control objective can be expressed in terms of the measured variables. A specific and rather general choice of the feature selection is proposed and validated on the illustrative example. In the current setting, a Random Forest regressor is used to fit the underlying static relationship but many other options are obviously possible. \\ \ \\ 
Undergoing work concerns the application of the framework on larger problems such as the Building Energy Management Systems where the absence of model is the main obstacle that faces the use of advanced MPC control in marketed solutions. 
\newpage 

\begin{figure}
\begin{center}
\includegraphics[width=\textwidth]{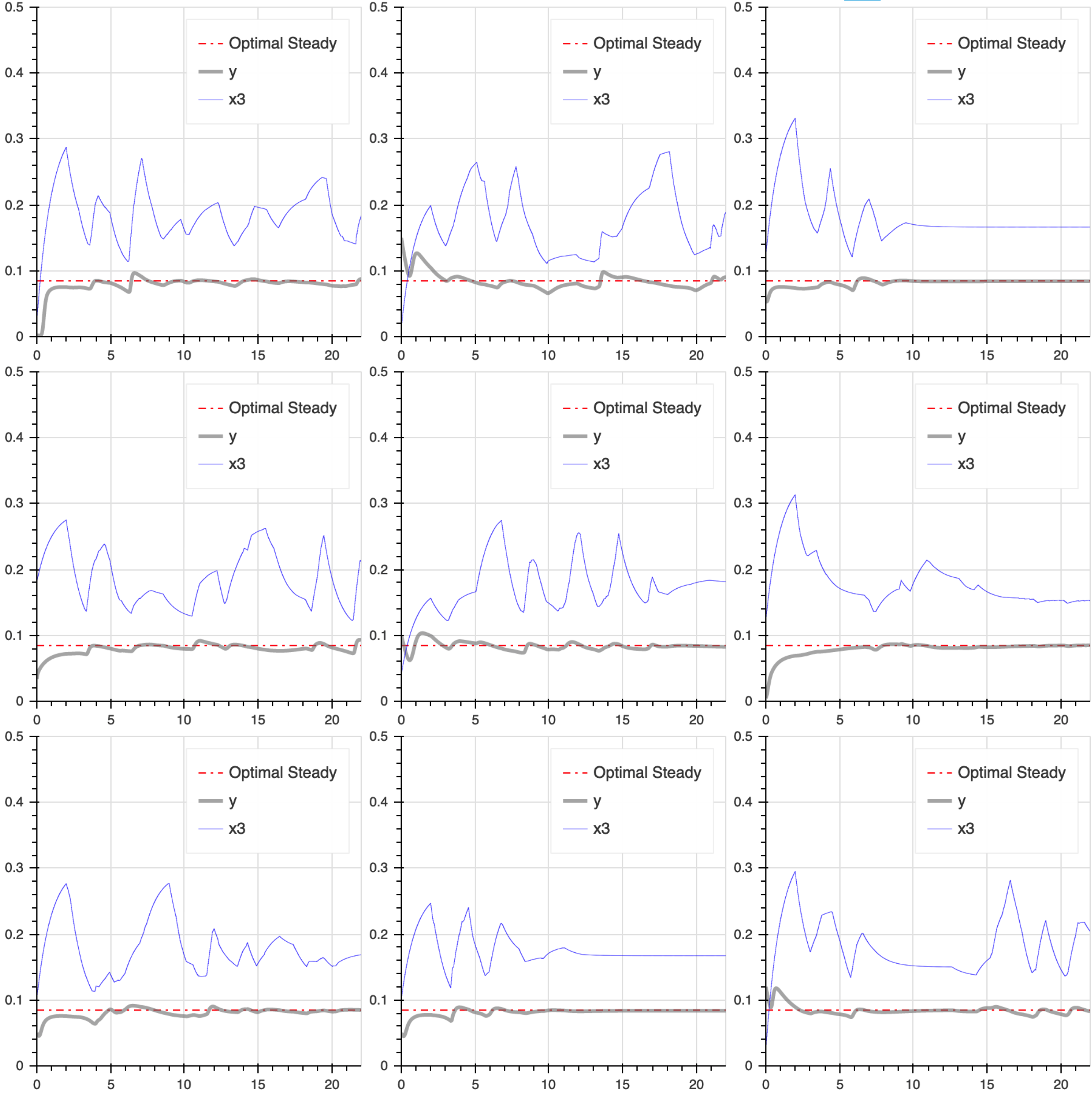}
\end{center}	
\caption{Evolution of the closed-loop system under the data-driven-based model predictive controller for different initial states. $N=100$, $N_{iter}=30$, $N_{guess}=1$, Two decision variables  at instants $\{0,10\}$.} \label{Res_N100_dof2}
\newpage 
\end{figure}
\begin{figure}
\begin{center}
\includegraphics[width=\textwidth]{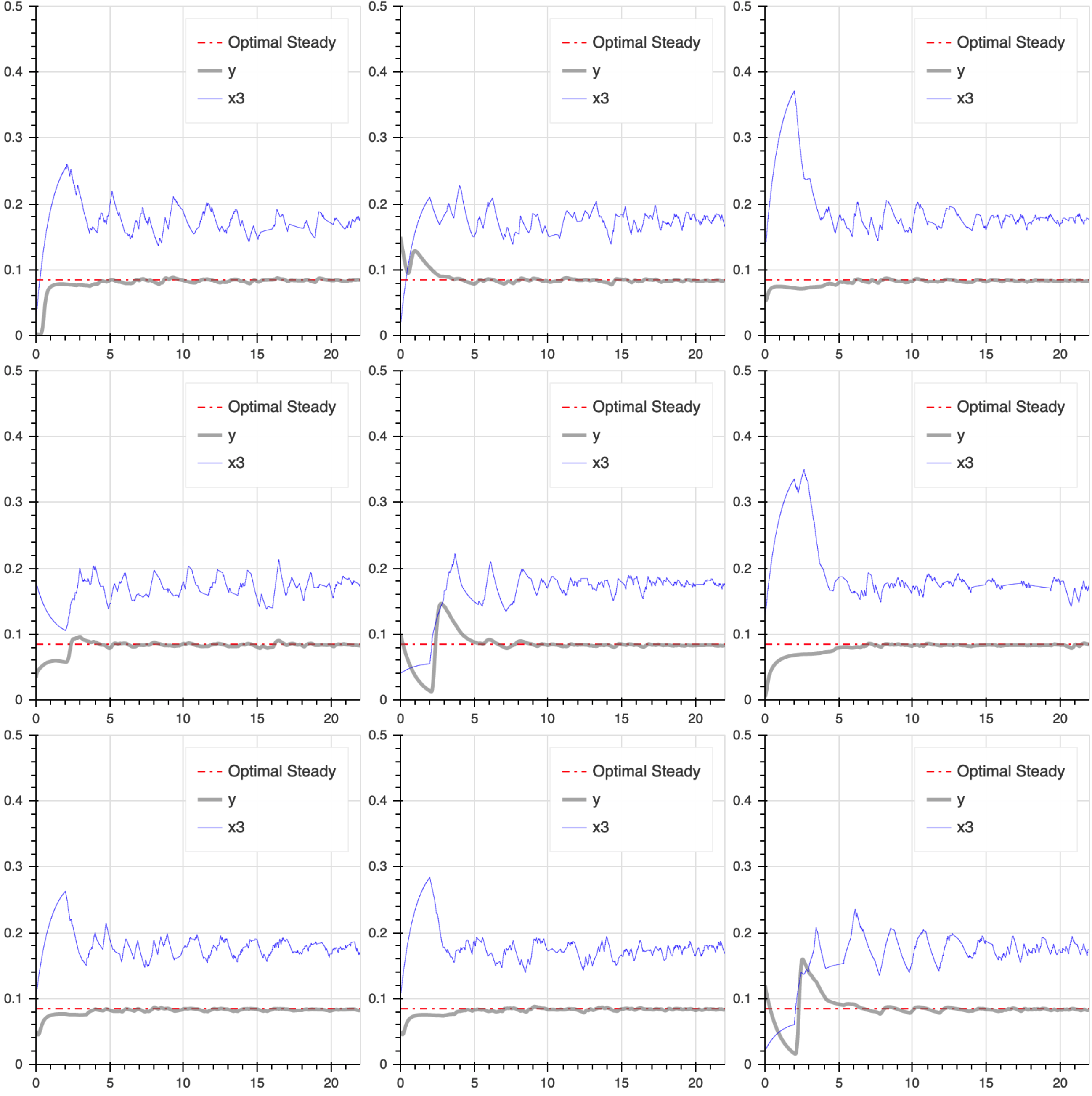}
\end{center}	
\caption{Evolution of the closed-loop system under the data-driven-based model predictive controller for different initial states. $N=100$, $N_{iter}=30$, $N_{guess}=1$, three decision variables  at instants $\{0,5,10,20\}$.} \label{Res_N100_dof4}
\end{figure}

\newpage 
\bibliographystyle{plain}
\bibliography{mabib.bib}

\end{document}